\documentclass{emulateapj}

\def\aap{\mbox{A\&A}}
\def\apjs{\mbox{ApJS}}

\def\ms{\mbox{$M_{\rm *}$}}
\def\mss{\mbox{$m_{\rm *}$}}

\def\mh{\mbox{$M_{\rm h}$}}

\def\phis{\mbox{$\Phi_s(\mss|\mh)$}}

\def\phisat{\mbox{$\phi_{\rm sat}$}}
\def\densub{\mbox{$\phi_{\rm sub}$}}
\def\phih{\mbox{$\phi_{\rm h}$}}

\def\NG{\mbox{$\langle N(>\ms|\mh)\rangle$}}

\def\Ns{\mbox{$\langle N_s(>\mss|\mh)\rangle$}}
\def\Nsat{\mbox{$\langle N_s\rangle$}}

\def\ng{\mbox{$n_g$}}

\def\xigg{\mbox{$\xi_{\rm gg}(r)$}}
\def\xiggh{\mbox{$\xi_{\rm gg}^{\rm 1h}(r)$}}
\def\xigghh{\mbox{$\xi_{\rm gg}^{\rm 2h}(r)$}}
\def\wp{\mbox{$w_{\rm p}(r_{\rm p})$}}

\def\msun{\mbox{M$_\odot$}}

\def\amt{\mbox{AMT}}
\def\shmr{\mbox{SHMR}}

\def\ssmr{\mbox{SSMR}}
\def\scmf{\mbox{subhCMF}}

\def\cmf{\mbox{CSMF}}  
\def\gsmf{\mbox{GSMF}}
\def\hmf{\mbox{HMF}}
\def\shmf{\mbox{subHMF}}

\def\msub{\mbox{$m_{\rm sub}$}}
\def\macc{\mbox{$m_{\rm sub}^{\rm acc}$}}
\def\mobs{\mbox{$m_{\rm sub}^{\rm obs}$}}

\def\lcdm{\mbox{$\Lambda$-CDM}}

\def\Nsub{\mbox{$\langle N_{\rm sub}(>\msub|\mh)\rangle$}}

\def\phisub{\mbox{$\Phi_{\rm sub}(\msub|\mh)$}}
\def\Psub{\mbox{$P_{\rm sat}(\mss|\msub)$}}
\def\Pcen{\mbox{$P_{\rm cen}(\ms|\mh)$}}
\def\phitwacc{\mbox{$\phi_{\rm sat,acc}$}}

\def\phigp{\mbox{$\phi_{\rm sat,obs}$}}

\def\Pacc{\mbox{$P_{\rm acc}(\mobs|\macc,z)$}}
\def\Psacc{\mbox{$P_{\rm sat}(\mss|\mobs)$}}
\def\Pobs{\mbox{$P_{\rm sat}(\mss|\mobs)$}}

\defcitealias{YMB09}{YMB09}
\defcitealias{Tinker-Wetzel11}{TW10}
\defcitealias{Giocoli08}{G08}
\defcitealias{Giocoli10}{G10}
\defcitealias{Angulo09}{A09}
\defcitealias{Gao04}{G04}
\defcitealias{Boylan-Kolchin10}{BK10}
\def\ltsima{$\; \buildrel < \over \sim \;$}    
\def\lesssim{\lower.5ex\hbox{\ltsima}}           
\def\gtsima{$\; \buildrel > \over \sim \;$}    
\def\grtsim{\lower.5ex\hbox{\gtsima}}           

\slugcomment{Submitted to ApJ}

\shorttitle{The satellite--subhalo mass relation}
\shortauthors{Rodriguez-Puebla, Drory \& Avila-Reese.}

\begin{document}


\title{The stellar--subhalo mass relation of satellite galaxies}


\author{A. Rodr\'iguez-Puebla, N. Drory and V. Avila-Reese}
\affil{Instituto de Astronom\'ia, 
Universidad Nacional Aut\'onoma de M\'exico,
A. P. 70-264, 04510, M\'exico, D.F., M\'exico.}

\email{apuebla@astro.unam.mx}



\begin{abstract}
  We extend the abundance matching technique (\amt) to infer the
  satellite--subhalo and central--halo mass relations (MRs) of local
  galaxies, as well as the corresponding satellite conditional mass
  functions.  We use the observed galaxy stellar mass function (\gsmf)
  decomposed into centrals and satellites and the \lcdm\
  distinct halo and subhalo mass functions as inputs. We explore the
  effects of defining the subhalo mass, \msub, at the time of
  (sub)halo accretion ($\macc$) versus defining it at the time of
  observation ($\mobs$); and we test the standard assumption that
  centrals and satellites follow the same MRs. We show that this
  assumption leads to predictions in disagreement with observations,
  specially when \mobs\ is used. Instead, we find that when the 
  satellite--subhalo MRs are constrained by the satellite \gsmf, 
  they are always different from the central--halo
  MR: the smaller the stellar mass, the less massive is the subhalo of
  satellites as compared to the halo of centrals of the same stellar
  mass. This difference is more dramatic when \mobs\ is used instead
  of \macc. On average, for stellar masses lower than $\sim 2\times
  10^{11}$ \msun, the dark mass of satellites decreased by $60-65\%$
  with respect to their masses at accretion time.  We find that MRs
  for both definitions of subhalo mass yield satellite conditional
  mass functions (\cmf) in agreement with observations. Also, when
  these MRs are used in a halo occupation model, the predicted
  two--point correlation functions at different stellar mass bins
  agree with observations.  The average stellar--halo MR is close to
  the MR of central galaxies alone, and conceptually this average MR
  is equivalent to abundance matching the cumulative total \gsmf\ to
  the halo + subhalo mass function (the standard \amt).  We 
  show that the use of \mobs\ leads to less uncertain MRs than \macc,
  and discuss some implications of the obtained satellite--subhalo MR.
  For example, we show that the tension between abundance and
  dynamics of Milky-Way satellites in the \lcdm\ cosmogony 
  gives if the faint-end slope of the \gsmf\ upturns
  to a value of $\sim-1.6$. 
\end{abstract}


\section{Introduction}
In recent years the abundance matching technique (AMT) has emerged as
a simple yet powerful statistical approach for connecting galaxies to
halos without requiring knowledge of the underlying
physics \citep[e.g.,] [and references therein]
{Vale+04,Kravtsov+04,Conroy+2006,Shankar+06,Baldry+2008,
  ConroyWechsler09,Drory+09,Behroozi+2010}.  

Briefly, the AMT assumes a one-to-one monotonic relationship between
stellar and halo masses which can be constrained by matching the
cumulative observed galaxy stellar mass function (\gsmf) to the
theoretical halo plus subhalo cumulative mass function.  Interestingly
enough, this simple approach successfully reproduces the observed
spatial clustering of galaxies
\citep[e.g.,][]{Conroy+2006,Moster+2010}. The AMT allows to probe the
average galaxy stellar--halo mass relation, \ms(\mh) (hereafter
\shmr), delivering very useful information for constraining models of
galaxy evolution \citep[e.g.,][]{Guo+10,Firmani+10,Avila-Reese+11}.

The above has motivated several authors to use the \amt\
extensively. For example, with the advent of large galaxy surveys at
different redshifts, the \amt\ has been applied for constraining the
evolution of the \emph{average} \shmr\ 
\citep[e.g.,][]{Conroy+2006,Drory+09,ConroyWechsler09,Moster+2010,Behroozi+2010}.
As a natural extension, these studies have been combined with
predicted average halo mass aggregation histories in order to infer
\emph{average} galaxy \ms\ growth histories as a function of mass
\citep[][see for a review Avila-Reese \& Firmani 2011, and references
therein]{ConroyWechsler09,FirmaniAvila10}.  By including observational
information on the gas content of galaxies, the \amt\ has been also
used to constrain the baryon mass to \mh\ relation of galaxies
\citep[][]{Baldry+2008,Rodriguez+2011}.  Finally, variants of the AMT,
where instead of mass functions, circular velocity functions or
functions of any other galaxy/halo global property are employed, have
been explored, too
\citep[e.g.,][]{Conroy+2006,Blanton+08,Trujillo-Gomez+11}.

The \amt\ has been commonly applied to the total (central plus
satellite galaxies) \gsmf\ matched against the total (distinct plus
satellite) halo population. This approach has been criticized, because
quite different average \shmr s are obtained for different proposed
forms of the satellite stellar-subhalo mass relation (\ssmr,
\mss(\msub)) and the central \shmr\ (\ms(\mh)\footnote{In order to
  make the distinction explicit, we shall use upper-case letters for
  the central galaxy and the distinct halo masses and lower-case
  letters for the satellite galaxy and subhalo masses.};
\citealp{Neistein11}).

A common (questionable) assumption is that the \ssmr\ is identical to
the central \shmr. Under this assumption, it is also common to 
define subhalo mass at the time of \emph{accretion} (\macc) rather than 
at the time of \emph{observation} (\mobs), when subhalos have lost a 
significant fraction of mass due to tidal stripping. The use of \macc\ has
been justified because this way is avoided the question of subhalo mass 
loss, and regarding the satellite \mss, it is expected that it remains almost 
constant since its infall into the host halo. The projected two-point 
correlation function of galaxies is reproduced under these assumptions 
\citep[][]{Conroy+2006,Moster+2010}. It should also be said that while the 
(local) \shmr\ for central galaxies has been determined 
\citep[e.g.,][]{Mandelbaum06,More11}, the stellar--subhalo mass relation 
for satellites/subhalos, \ssmr, has not been yet discussed in detail in
the literature.

In view of the above mentioned, some important questions arise. 
Why does using \macc\ instead
of \mobs\ lead to the correct clustering of galaxies? Does the
\macc--\ms\ relation reproduce the observed satellite \gsmf, the
conditional stellar mass function, and spatial clustering of galaxies
at the same time? Even more fundamentally, if is not assumed that the \ssmr\ is
identical to the central \shmr, then, what follows for the \ssmr,
either using \mobs\ or \macc?  Does it deviate from the central
\shmr?

In this paper we extend the common \amt\ to constrain both the central
\shmr\ and the \ssmr\ separately, as well as the average (total)
\shmr. By construction, this formalism also allows to predict the mean
satellite conditional mass function (\cmf), i.e., the probability that
satellites of a given stellar mass reside in distinct host halos of a
given mass. We will (i) test whether the \ssmr\ and the central \shmr\
follow the same shape; (ii) discuss the consequences of defining the
subhalo mass at accretion time vs.\ at observed (present) time; and (iii) check
the self-consistency of our predicted present-day central \shmr\ and
\ssmr\ by comparison with the observed satellite \cmf\ and the spatial
clustering of galaxies.

This paper is laid out as follows. In Section 2 we present the \amt,
focusing on the details of  our extended abundance matching.  
In Section 3 we present the predicted
stellar-halo mass relations (\S\S 3.1) and satellite $\cmf$s (\S\S 3.2)
for cases when the \ssmr\ is assumed equal to the central \shmr, and
when both mass relations are independently constrained. In \S\S 3.3, a
Halo Occupation Distribution (HOD) model is used to explore whether
the predicted central \shmr's and \ssmr's are consistent with the
observed spatial clustering of galaxies.  Section 4 is devoted to our
conclusions and a discussion of the results and their implications.

All our calculations are based on a flat $\Lambda$CDM cosmology with
$\Omega_\Lambda=0.73$, $h=0.7$, and $\sigma_8=0.84$.

\section{The Abundance Matching Technique}

In this section we describe the technique of matching abundances
between central galaxies and halos and satellite galaxies and subhalos, 
separetely, which we present here as an extension to the standard \amt.

\subsection{Modeling the central \& satellite $\gsmf$s}

To model the central \gsmf, let $\Pcen$ denotes 
the probability distribution function that a distinct halo of mass \mh\ hosts
a central galaxy of stellar mass  $\ms$.  
Then the number density of central galaxies
with stellar masses between $\ms$ and $\ms+ d\ms$ is given by
\begin{equation}
\phi_{\rm cen}(\ms)d\ms=d\ms\int^{\infty}_{0}\Pcen\phih(\mh)d\mh.
\label{Pcen}
\end{equation}
 
For the population of satellite galaxies in individual subhalos,
let \Psub\ be the probability distribution function 
that a subhalo \msub\footnote{Whenever
we use \msub\ we refer to subhalo mass generically. In practice,
that can either be the mass at accretion time, \macc, or at
observation (present-day) time, \mobs.} hosts a satellite galaxy of
stellar mass  $\mss$. 
Thus the average satellite \cmf\ (the number of
satellite galaxies of stellar mass between $\mss$ and $\mss+ d\mss$
that reside in distinct host halos of mass \mh, e.g., \citealp*{Yang+2009}) is
\begin{equation}
\phis d\mss=d\mss\int^{\infty}_{0}\Psub\phisub d\msub,
\label{cmfsat}
\end{equation}
where \phisub\ is the subhalo conditional mass function \citep[\scmf, i.e., the
number of subhalos of mass between $\msub$ and $\msub+ d\msub$ residing in host halos of
mass \mh; e.g.,][]{Boylan-Kolchin10}. A natural 
link between the satellite \gsmf, \phisat,
and the distinct halo mass function (\hmf, \phih) arises 
once the satellite \cmf\ is given:
\begin{equation}
\phisat(\mss)d\mss=d\mss\int^{\infty}_{0}\phis\phih(\mh) d\mh.
\label{densat}
\end{equation}
Inserting equation (\ref{cmfsat}) into equation (\ref{densat}) and
rearranging terms, the satellite \gsmf\ can be rewritten in terms of
\Psub:
\begin{equation}
\phisat(\mss)d\mss=d\mss\int^{\infty}_{0}\Psub\densub(\msub)d\msub,
\label{phisat}
\end{equation}
where the subhalo mass function (\shmf) is given by
\begin{equation}
\densub(\msub)d\msub=d\msub\int^{\infty}_{0}\phisub\phih(\mh)d\mh.
\label{den_sub}
\end{equation}
Equations (\ref{Pcen}) and (\ref{phisat}) describe the abundance matching
in its differential form for the central-halo and  satellite-subhalo populations, 
respectively. The distribution probability \Pcen\ is defined by
the mean \ms(\mh) relation and a scatter around it of $\sigma_c$,
while the distribution probability
\Psub, assumed to be independent of host halo mass, is
defined by the mean \mss(\msub) relation and a scatter around it
of $\sigma_s$. Observe that once
\Psub\ is given, the satellite \cmf\ is a prediction according to
equation (\ref{cmfsat}).

Here, \Pcen\ and \Psub\ are modeled as lognormal distributions with
a width (scatter around the stellar mass) assumed to be constant and the
same for both centrals and satellites,  $\sigma_c=\sigma_s=0.173$~dex. 
Such a value was inferred for central galaxies 
 from the analysis of general large group catalogs
(\citetalias{YMB09}) and it is supported by
recent studies on the kinematics of satellite galaxies  \citep{More11}.
Regarding the intrinsic scatter of the satellite-subhalo relation, it has 
not been discussed in detail in the literature. While the exploration of this
scatter is beyond the scope of the present paper, our conclusions
will not depend critically on the assumed value for it or
even if it is allowed to depend on host halo mass.  We will further
discuss this question in \S4.2. Both \mss(\msub) and \ms(\mh) are
parametrized by the same modified two-power-law form proposed in
\citet{Behroozi+2010}.  This five-parameters function is quite general
and, in the context of the \amt, has been shown to reproduce the main
features of a Schechter-like \gsmf.

\subsection{The relation to standard abundance matching}

In the standard \amt\ the cumulative halo+subhalo mass function and the total 
observed cumulative \gsmf\ are matched to
determine the mass relation between halos and galaxies,
which is assumed to be monotonic. In this context, 
no intrinsic scatter in the stellar mass at a given
halo is assumed. In our approach, where the galaxy and halo populations are 
separated into centrals/satellites and distinct halo/subhalos, the latter entails that 
the probability distribution functions of centrals and satellites take the particular 
forms: $P_{\rm cen}(\mathcal{M}|\mh)=\delta(\mathcal{M}-\ms(\mh))$ and
$P_{\rm sat}(\mathcal{M}|\msub)=\delta(\mathcal{M}-\mss(\msub))$,
where \ms(\mh) and \mss(\msub) are the mean central-halo and satellite-subhalo mass
relations, and $\delta$ is for the $\delta$-Dirac function. The above "no scatter''
probability distribution function for centrals applied in Eq. ($\ref{Pcen}$) 
would lead us to conclude that the cumulative central \gsmf, $n_{\rm cen}(>\ms)$, 
should match the cumulative distinct halo mass function, $n_{\rm h}(>\mh(\ms))$. 
The same reasoning applies for satellites/subhalos. 
Therefore, we arrive to the standard \amt\ formulation:
\begin{equation}
\ng(>\ms)=n_{h}(>M_{\rm h})+n_{\rm sub}(>M_{\rm h}),
\end{equation}
where $\ng(>\ms)\equiv n_{\rm cen}(>\ms)+n_{\rm sat}(>\ms)$ is the
total \gsmf. 

Since the abundance matching can be applied to centrals/halos
and satellites/subhalos separately, let analyze now only the 
latter. Under the assumption that the \mss(\msub) relation
is independent of the host halo mass, it is clear that using either the
abundance matching of all satellites and all subhalos, 
$n_{\rm sat}(>\mss)=n_{\rm sub}(>\msub)$, or the
matching of their corresponding mean occupational numbers,
one may find exactly the same \mss(\msub) relation. In this
sense, we state that matching abundances is equivalent
to matching occupational numbers:
\begin{eqnarray}
\Ns=\Nsub & \\ 
\Longleftrightarrow \nonumber
n_{\rm sat}(>\mss)=n_{\rm sub}(>\msub).
\end{eqnarray}

In the case that the probability distribution function \Psub\
includes scatter around the mean \ssmr, 
as we consider here, the above conclusion remains the same 
whilst \Psub\ is assumed to be independent on halo mass. 
In general, the inclusion of constant scatter in the galaxy-halo mass
relations is not a conceptual problem for the \amt,
but it slightly modifies the shape of the mass relations at the
high mass end \citep[see][]{Behroozi+2010}. 
Finally, note that if \Psub\ depends on \mh, then $\phisat$ may not be directly 
related to $\densub$ (see Eq. 4) and using either the matching of satellites and
subhalo abundances or the matching of their
corresponding occupational numbers would not lead to find exactly
the same \mss(\msub).

\subsection{Inputs for matching abundances}

The inputs required for the procedure described above are the
\scmf, the distinct \hmf, and the observed satellite and central
$\gsmf$s.

For the \scmf, we use the results obtained in
\citet[][\citetalias{Boylan-Kolchin10}]{Boylan-Kolchin10} based on the
analysis of the Millennium-II Simulation. This is a re-simulation with
the same resolution of a smaller volume of the Millennium
Simulation. It consists of $2160^3$ particles, each of mass
$m=6.885\times 10^6h^{-1}\msun$ in a periodic cube of length
$L=100h^{-1}$Mpc. Observe that this mass particle resolution is around
four orders of magnitude below the subhalo masses required
($\sim10^{10}h^{-1}\msun$) to match the lower stellar mass limit in
the \citetalias{YMB09} \gsmf.
The fitting formula for the cumulative \scmf\ reported in
\citetalias{Boylan-Kolchin10} at  the 
$[10^{12},10^{12.5}]h^{-1}\msun$ mass interval is:
\begin{equation}
\Nsub=\mu_0\left(\frac{\mu}{\mu_1}\right)^a\exp\left[-\left(\frac{\mu}{\mu_{\rm cut}}\right)^b\right],
\label{cmf_sub}
\end{equation}
where $\mu=\msub/\mh$.  For $\msub$$=\macc$,  $(\mu_0,\mu_1,\mu_{\rm
cut},a,b,)=(1,0.038,0.225,-0.935,0.75)$, while for  $\msub=\mobs$, $=(1.15^{(\log M_{\rm h}-12.25)},
0.01,0.096,-0.935,1.29)$. According to \citetalias{Boylan-Kolchin10}, 
the shape of the \mobs\ \scmf\ remains the same for other halo masses
but its normalization, $\mu_0$, systematically increases with \mh, roughly 
by 15\% per dex in \mh. Such a behavior has been reported 
in an analysis of the Millennium simulations by 
\citet[][for closely related results,   with small differences in the amplitude,
see also \citealp{Gao04, vandenBosch05,Zentner+2005,Angulo09,Giocoli10}]{Gao+2011}.
we introduce the dependence $\mu_0=1.15^{(\log M_{\rm h}-12.25)}$, 
where $\mu_0=1$ at $\log\mh=12.25$.
In the case of the \macc\ \scmf, the normalization factor is nearly independent on 
\mh, i.e., $\mu_0=1$ (\citetalias{Boylan-Kolchin10}; see also \citealp{Giocoli08}).
The \scmf\ is given by $\Phi_{\rm sub}=dN_{\rm sub}/d\msub$.

In order to construct the \macc\ \scmf, \citetalias{Boylan-Kolchin10}
traced each subhalo back in time so that they were able to find the
point at which its bound mass reached a maximum, i.e., the time the
halo became a subhalo. The latter guarantees that we are working with
the surviving population of accreted halos and no further assumptions
on the merging process are necessary.

The difference between the Millennium-II simulation cosmology  and
ours leads to differences in the resulting abundances of
subhalos of roughly a few percent in the amplitude of the subhalo
mass function (\citetalias{Boylan-Kolchin10}). This
is also supported by previous works that explored the impact of
changing cosmological parameters on the subhalo occupational
statistics (e.g., \citealp{Zetner+2003}). 
Additionally, to be consistent with the same cosmology for which the subhalo \scmf\ 
was inferred, we repeated all the analysis to be showed below but using the WMAP1
cosmology.  We find that all our results are practically the same.

For the distinct \hmf, we will use the formula given by
\citet{Sheth-Tormen99}.  This formula provides a reasonable fit to the
the virial mass\footnote{The mass enclosed within the radius at which,
  according to the spherical collapse model, the overdensity of a
  sphere is $\Delta_{\rm vir}$ times larger than the \textit{matter}
  critical density of the used cosmological model; for the cosmology
  assumed here, $\Delta_{\rm vir}$($z=0$)$=97$.}  function at $z\sim
0$ measured in large cosmological N-body simulations
\citep[e.g.,][]{Klypin+11,Cuesta08}.

For our purpose, the decomposition of the \gsmf\ and the $\cmf$s into
centrals and satellites galaxies is necessary. Using a large general
group catalog \citep{Yang+07} based on the data from the SDSS,
\citetalias{YMB09} constructed and studied the decomposition of the
\gsmf\ and the $\cmf$s into centrals and satellites galaxies. In that
paper, a central galaxy was defined as the most massive galaxy in a
group and the remaining galaxies as satellites.  For the mass
completeness limit in the \gsmf, they adopted the value as function of
redshift proposed in \citet{vandenBosch+08}. They have also taken into
account incompleteness in the group mass by considering an empirical
halo-mass completeness limit (for details we refer the reader to
\citetalias{YMB09}).

Where necessary, halo masses are converted to match our definition of
virial mass and stellar masses are converted to the
\citet{Chabrier03} IMF. In particular, \citetalias{YMB09} defined
halo masses as being 180 times the background density, according to
\citet[][see their appendix B]{Giocoli10} these halos are $\sim11\%$ larger than
our definition of virial mass. 

\subsection{Procedure and uncertainties}

We constrain the parameters of the functions proposed to describe
the central \shmr\ and \ssmr\ by means of Eqs. (\ref{Pcen}) and (\ref{densat}),
and by using the Powell's directions set method in multi-dimensions
for the minimization  \citep{Press92}. Note that in our analysis the 
reported statistical errors in the \gsmf s, as well as 
the intrinsic scatter in the mass relations are taken into account. 
However, we will not analyze rigorously here the effects of 
uncertainties on the mass relations as well as their errors. 
Instead, we remit the reader to previous works \citep{Moster+2010,Behroozi+2010,
Rodriguez+2011}.

\citet{Behroozi+2010} studied in detail the uncertainties and effects on the average \shmr\ 
due to different sources of error like those in the observed \gsmf s, including
stellar mass estimates; in the halo mass functions; 
in the  uncertainty of the cosmological parameters; 
and in linking galaxies
to halos, including the intrinsic scatter in this connection. These
authors have found that the largest uncertainty by far in the
\shmr\ is due to systematic shifts in the stellar estimates. 
The second important source of uncertainty is due to the
intrinsic scatter, that we take into account in our analysis.
Other statistical and sample variance errors have negligible
effects, at least for local galaxies. According to the 
\citet[][]{Behroozi+2010} study, the statistical and systematical uncertainties
account for 1$\sigma$ errors in the \shmr\ of approximately 0.25 dex at
all masses, which is almost totally due to the uncertainty in stellar mass estimates. 
We have explored here also the effects of the subhalo CMF uncertainty on the
SSMR. By using the 25\% per dex in \mh\ variation reported by \citet{Giocoli10}
(instead of 15\%), we find that the SSMR shifts in \mss\ by only $\approx 0.04$ dex.

\begin{figure}
\vspace*{-240pt}
\includegraphics[height=6.5in,width=6.5in]{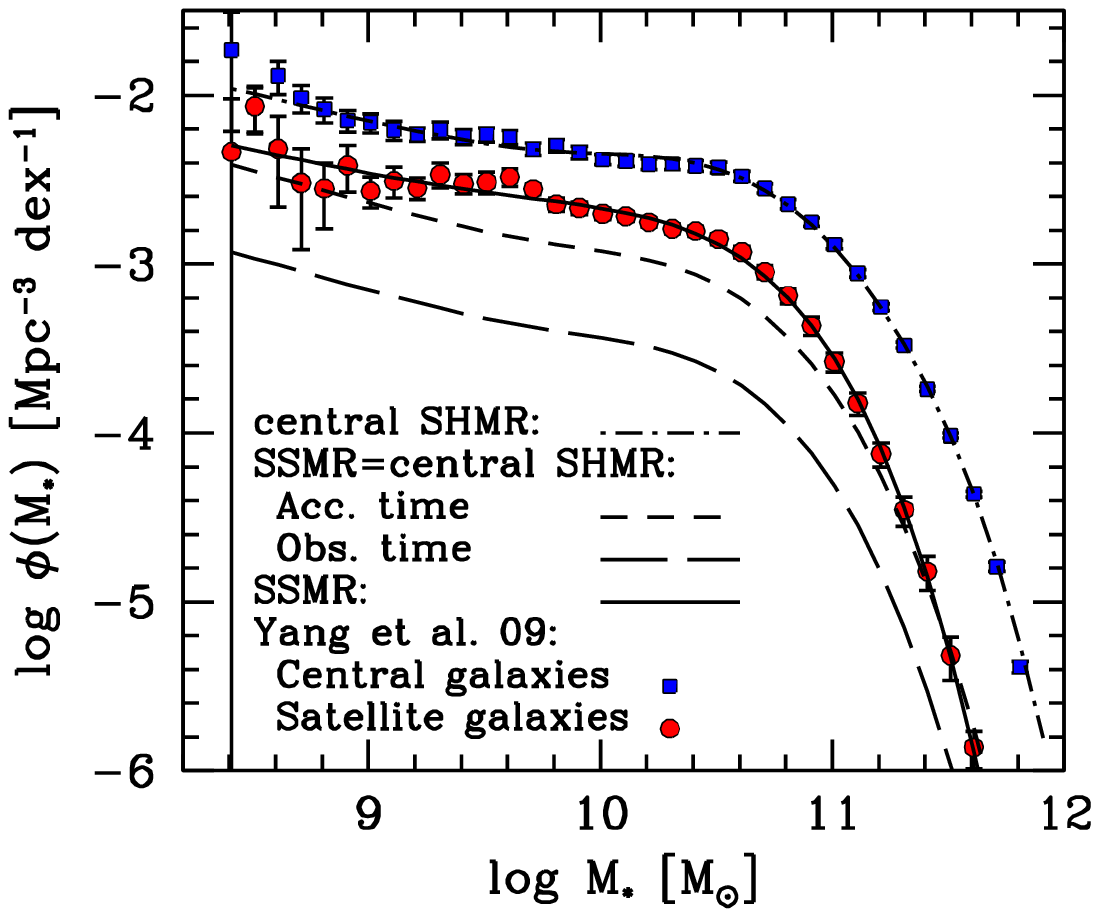}
\caption{Satellite $\gsmf$s calculated under the assumption 
that $\Psub=\Pcen$ and for the cases \macc\ (long-dashed line)
and \mobs\ (short-dashed line) were used for the subhalo mass
definition. Filled circles and squares with error bars show the
\citetalias{YMB09} central and satellite
$\gsmf$s, respectively. The solid line is for the case when 
\Psub\ was determined using the \citetalias{YMB09} satellite \gsmf\ as a constraint,
i.e., is the best model fit to this function.
}
\label{gsmf}
\end{figure}

\section{Results}

\begin{figure*}
\vspace*{-240pt}
\includegraphics[height=6.5in,width=6.5in]{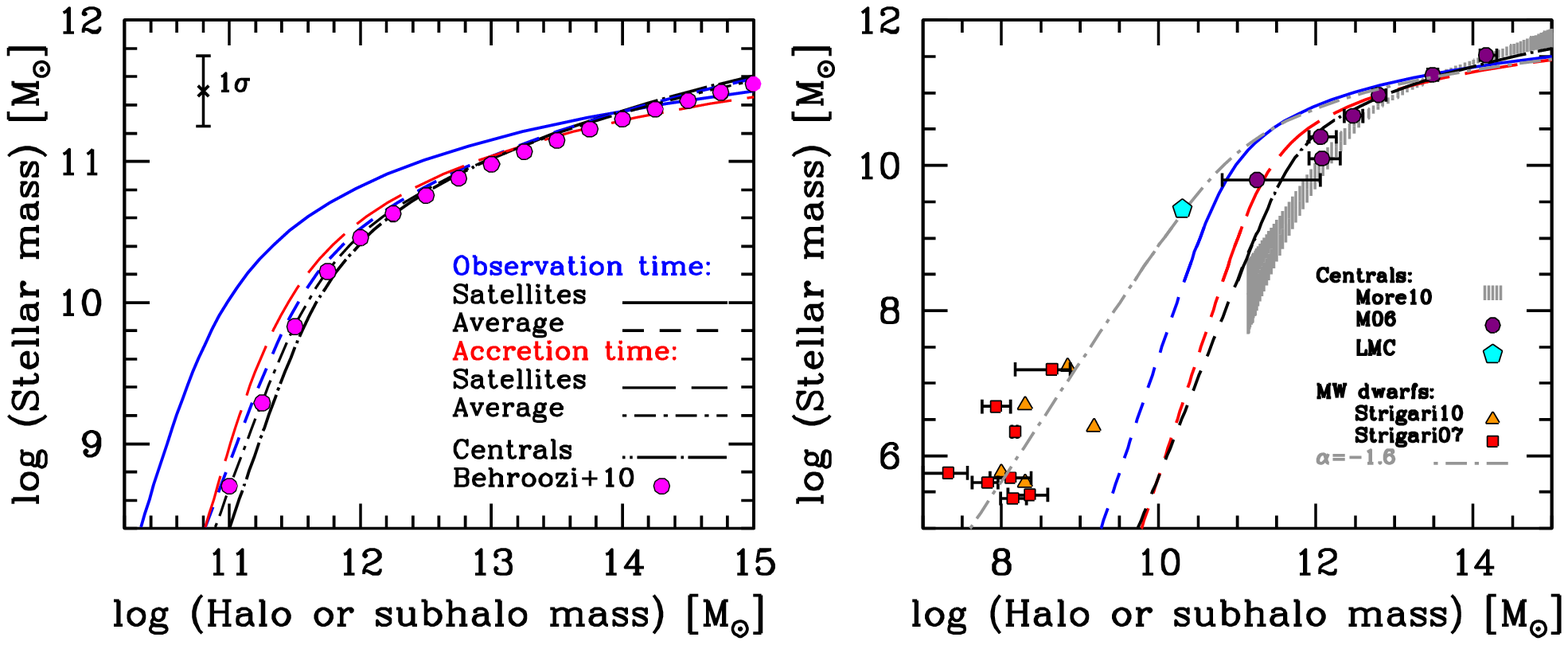}
\caption{ \textit{Left panel:} 
Inferred mass relations for satellite galaxies when the subhalo 
mass is defined as \mobs\ (blue solid line) and as \macc\ (red long-dashed-line),
and for central galaxies (black dot-long-dashed line). 
The density-weighted average relation when using subhalo mass \mobs\ 
is plotted with a blue short-dashed line, while when using subhalo mass \macc\ is plotted with 
a gray short-dotted line. For comparison, the (average) mass relation obtained in 
\citet{Behroozi+2010} is also plotted (filled circles).
\textit{Right panel:} A comparison of the mass relations of satellite and central 
galaxies with direct observational inferences (the same line code of left panel for
models is used; dashed lines indicate extrapolations to lower masses). Filled circles with 
error bars correspond to the mass relation of central galaxies from the analysis of 
staked weak-lensing in \citet{Mandelbaum06}. Dashed area indicates the 68\% 
of confidence in the mass relation of central galaxies
using the kinematics of satellites \citep{More11}. 
The inferred total mass at the tidal radii for the brightest dwarf galaxies obtained in
 \citet{Strigari+2007} and  \citet{Strigari+2010} are plotted with filled squares
 and triangles, respectively. Filled pentagon shows the mass at the tidal
 radius for the Large-Magallanic Cloud \citep{Weinberg1999}.
 The gray dotted-dashed curve is the \mobs\ \ssmr\
 assuming a faint-end slope in the satellite \gsmf\ of $\alpha=-1.6$.}
\label{msmh}
\end{figure*}

\subsection{The satellite \gsmf\ and the \ssmr}

By means of the procedure described in Section 2, we calculate first
the satellite \gsmf\ (Fig.~\ref{gsmf}) when the \ssmr\ and the central
\shmr\ are assumed to be the same, i.e., $\mss(\msub)=\ms(\mh)$.  This
is equivalent to assume that $\Psub=\Pcen$ if the intrinsic scatter of
both relations is the same.  We obtain the central \shmr\ by matching
abundances of YMB09 central galaxies to distinct halos. This relation
and the subhalo mass function obtained from the theoretical \scmf\
(eq. \ref{cmf_sub}), are used to infer the satellite \gsmf\
(eq. \ref{phisat}).  The satellite \gsmf\ is presented for the two
cases of subhalo mass definition: \phitwacc\ when \macc\ is used
(long-dashed line), and \phigp\ when \mobs\ is used (short-dashed
line). The observational results of \citetalias{YMB09} are
plotted as well.

Under the assumption that $\mss(\msub)=\ms(\mh)$, the predicted number
density of satellites at masses below the knee is underestimated on
the average by a factor of $\sim 2$ when using \macc, and $\sim 5$
when using \mobs.  Note that the former is closer to the
\citetalias{YMB09} data.  The reason is simply because the
normalization of the \macc\ \shmf\ is higher and closer to the
distinct \hmf\ than the normalization of the \mobs\ \shmf.  Therefore,
satellites of equal \mss\ are expected to have a higher number density
when using the accreted-time (\macc) \shmf\ compared to using the
observed-time (\mobs; present-day) \shmf.

However, neither \macc\ nor \mobs\ are able to reproduce the observed
satellite \gsmf, and the discrepancy is due to the basic assumption of
a common stellar mass--(sub)halo mass relation for centrals and
satellites. In the case that $\msub=\macc$, this is equivalent to assume that the \ssmr\ is 
independent of redshift. But in fact this cannot be the case since the satellite mass \mss\ 
hardly will remain the same since it was accreted to the present epoch. On the other hand, 
when using $\msub=\mobs$, that the \ssmr\ is equal to the central \shmr\ implies that both have
evolved, on average, identically. This cannot be the case because it is evident that
the population of subhalos evolved differently to distinct halos, mainly by losing mass 
due to tidal striping \citep[e.g.,][]{Kravtsov+04,vandenBosch05}. 

The next step in our analysis is to allow the \ssmr\ and central
\shmr\ to be different, i.e., $\mss(\msub)\ne\ms(\mh)$. In this case, \Psub\
is determined by means of Eq. (\ref{phisat}) using the
\citetalias{YMB09} satellite \gsmf\ as a constraint. For the \scmf, we
again use both definitions of subhalo mass, \macc\ and \mobs. For
illustrative purpose, we present the resulting satellite \gsmf\ for
the case when \mobs\ was used (solid line in Fig~\ref{gsmf}; an almost
identical \gsmf\ is obtained when \macc\ was used).

As shown in Fig. ~\ref{msmh}, the \ssmr s obtained by using \macc\
(long-dashed line) and \mobs\ (solid line) are quite different.
The central \shmr\ (dot-dashed line) is the same for both cases.
The error bar in the left panel shows a 1$\sigma$ uncertainty
of 0.25 dex in the normalization of the mass relations. This is roughly
the uncertainty estimated by \citet{Behroozi+2010} taking into account
all the systematical and statistical sources of errors (see \S\S2.4). 

When using the accretion-time subhalo mass, \macc, we note that the 
resulting \ssmr\ at $\log(\mss/\msun)<11$ systematically lies above
the central \shmr, with differences in the stellar-mass axis (halo-mass axis) 
of $\sim0.5$ dex (0.2 dex) at the smallest masses. For $\log(\ms/\msun)>11$ 
this trend is inverted, but the differences between central and satellites are 
very small. However, the relation
obtained this way should be taken with caution. By construction, each
\macc\ is itself a cumulative distribution of all the objects accreted
in a time interval $\Delta z$. Therefore such a \ssmr\ entails that
all accreted objects of mass \macc\ would evolve, on average, to host the same \mss\
despite having been accreted at \emph{different times}. We discuss this in \S\S 4.1.

When using the observation-time (present-day) subhalo mass, \mobs, the
\ssmr\ (solid blue line) and central \shmr\ are very different,
though they show the same trend as when using \macc.  For example, on
average, a satellite with $\log(\ms/\msun)=10$ resides in a subhalo a
factor of $\sim4$ less massive than the halo of a central galaxy with
the same stellar mass.  Notice that $\mobs(\mss)<\macc(\mss)$ and that
the difference increases the lower the mass is.  This is consistent
with the picture that most massive subhalos, on average, fell into
larger halos just very recently and they have not had time to lose
significant amounts of mass due to tidal striping, in contrast to the
lowest mass subhalos.

This also suggests that the \ssmr\ for both definitions of subhalo
mass should tend to the central \shmr\ at the high-mass end, but this
is not the case as seen in Fig. ~\ref{msmh} where small differences
remain.  The possible reasons are that, firstly, the intrinsic
scatter around the stellar--(sub)halo mass relations is actually lower
for the former than for the latter (here we assumed it to be the same
for satellites and centrals, see \S4.2). Secondly, that the \citetalias{YMB09}
satellite \gsmf\ may underestimate the true satellite mass function at
large masses \citep[see also][]{Skibba+2011}.

Fiber collisions could introduce some systematic error that may affect the 
\citetalias{YMB09} group catalog. To study the impact of this possible systematic 
error, \citetalias{YMB09} divided their group catalog into two sample: one that 
uses galaxies with known redshifts, and another that includes galaxies that 
lack redshifts due to fiber collisions. When compared the corresponding satellite
$\cmf$s from both samples (see their Fig. 6), they found that the sample for which
the correction for fiber collisions has been taking into account, has a higher 
amplitude of the $\cmf$s than when this correction has not been applied, specially 
in low mass halos.  However, the difference is very marginal and well within the 
error bars. We conclude that fiber collisions in the \citetalias{YMB09} group
catalog are not a serious source of systematics able to affect our conclusions.
Regarding completeness and contamination of their group catalog \citep[for
details see][]{Yang+07}, 80\% 
 have a completeness greater than 0.6, while 85\% have a contamination lower than
0.5. In terms of purity, their halo-based group finder is consistent with the ideal situation.

Finally, we note that the mass relation usually obtained by matching
abundances between the {\em total} \gsmf\ and the halo plus subhalo
mass function, in the light of the decomposition into centrals and
satellites, could be interpreted as a \emph{density--weighted average}
\shmr:
\begin{equation}
\langle\ms(M)\rangle_{\phi}=\frac{\densub(M)}{\phi_{\rm DM}(M)}\mss(M)+\frac{\phih(M)}{\phi_{\rm DM}(M)}\ms(M),
\end{equation}
where $\phi_{\rm DM}(M)=\densub(M)+\phih(M)$, $\mss(M)$ is the mean
\ssmr\ and $\ms(M)$ is the mean central \shmr. This relation is
plotted in Fig. \ref{msmh} with short-dashed-dot and
short-dashed lines when using \macc\ and \mobs, respectively. Since most
galaxies in the YMB09 catalog are centrals, the central $\shmr$ is
very close to the density-weighted average \shmr.  For comparison, we
plotted the \citet{Behroozi+2010} average mass relation (filled
circles), which is in excellent agreement with our density-weighted
average \shmr\ when using the accreted-time subhalo mass, \macc.

Observe that differences between the satellite and the average (total)
mass relations are small when \macc\ is used, while differences become
dramatic when \mobs\ is used.  The above explains why under the
assumption that $\mss(\macc) = \ms(\mh)=\langle\ms(M)\rangle_{\phi}$, the
resulting satellite \gsmf\ are closer to observations.  On the other
hand, since the \macc\ \shmf\ has a higher normalization than the
\mobs\ \shmf, the above shows that when assuming $\mss(\mobs) =
\ms(\mh)=\langle\ms(M)\rangle_{\phi}$, we should expect that the resulting satellite \gsmf\
is significantly below the observed satellite \gsmf.

\subsubsection{Comparison with other observational inferences}

In the right panel of Fig.~\ref{msmh}, we plot some observational
inferences of halo and subhalo masses as a function of stellar mass. 
The inferred $\langle\mh\rangle(\ms)$ of central galaxies from staked weak-lensing 
studies using the SDSS \citep{Mandelbaum06} are shown as filled circles 
with error bars. \citet{Mandelbaum06} reported the data actually for 
blue and red galaxies separately. 
We estimated the average mass relation for central galaxies as:
$\langle\mh\rangle(\ms)=f_b(\ms)\langle\mh\rangle_b(\ms)+
f_r(\ms)\langle\mh\rangle_r(\ms)$, where $f_b(\ms)$ and $f_r(\ms)$ are 
the blue and red galaxy fractions in the sample, and 
$\langle\mh\rangle_b$ and $\langle\mh\rangle_r$ are the 
corresponding blue and red mass relations. 
The inferred $\langle\log(\ms)\rangle(\mh)$ for central galaxies 
from staked kinematics of satellites \citep{More11} are plotted as the
dashed area indicating the 68\% of confidence. 

Our inferred central \shmr\ (dotted-dashed curve) is consistent with 
the weak-lensing inferences at all masses, and with the satellite 
kinematics inferences at masses $\ms\grtsim 10^{11}$ \msun;
for smaller masses, our halo masses are a factor up to $\sim 2$ 
smaller than the satellite kinematics inferences. In fact, it was
already noted that using the kinematics of satellite galaxies yields halo masses around
low mass galaxies that are systematically larger than most other 
methods, specially for red central galaxies \citep{More11,Skibba+2011,Rodriguez+2011}.

Regarding satellites, unfortunately, there are not direct inferences of
their subhalo masses. Some model-dependent estimates based on
dynamical observations of Milky-Way (MW) satellites were presented in the
literature. For example, using the line-of-sight
velocity dispersions measured for the brightest spheroidal 
dwarf galaxies, \citet{Strigari+2007} and \citet{Strigari+2010} determined their 
masses within their tidal radii. These dynamical masses, plotted
in Fig.~\ref{msmh} (filled squares and triangles, respectively), are expected
to be of the order of \mobs. We also plot an estimate of the mass at the
tidal radius for the Large-Magallanic Cloud \citep[filled pentagon,][]{Weinberg1999}. 
The \ssmr s constrained here do not extend to the small masses
of MW satellites but we plot their extrapolations to these masses (dashed
curves). The gray dotted-dashed curve will be discussed in \S\S 4.3

\subsection{The satellite \cmf}

\begin{figure*}
\vspace*{-200pt}
\includegraphics[height=6.5in,width=6.5in]{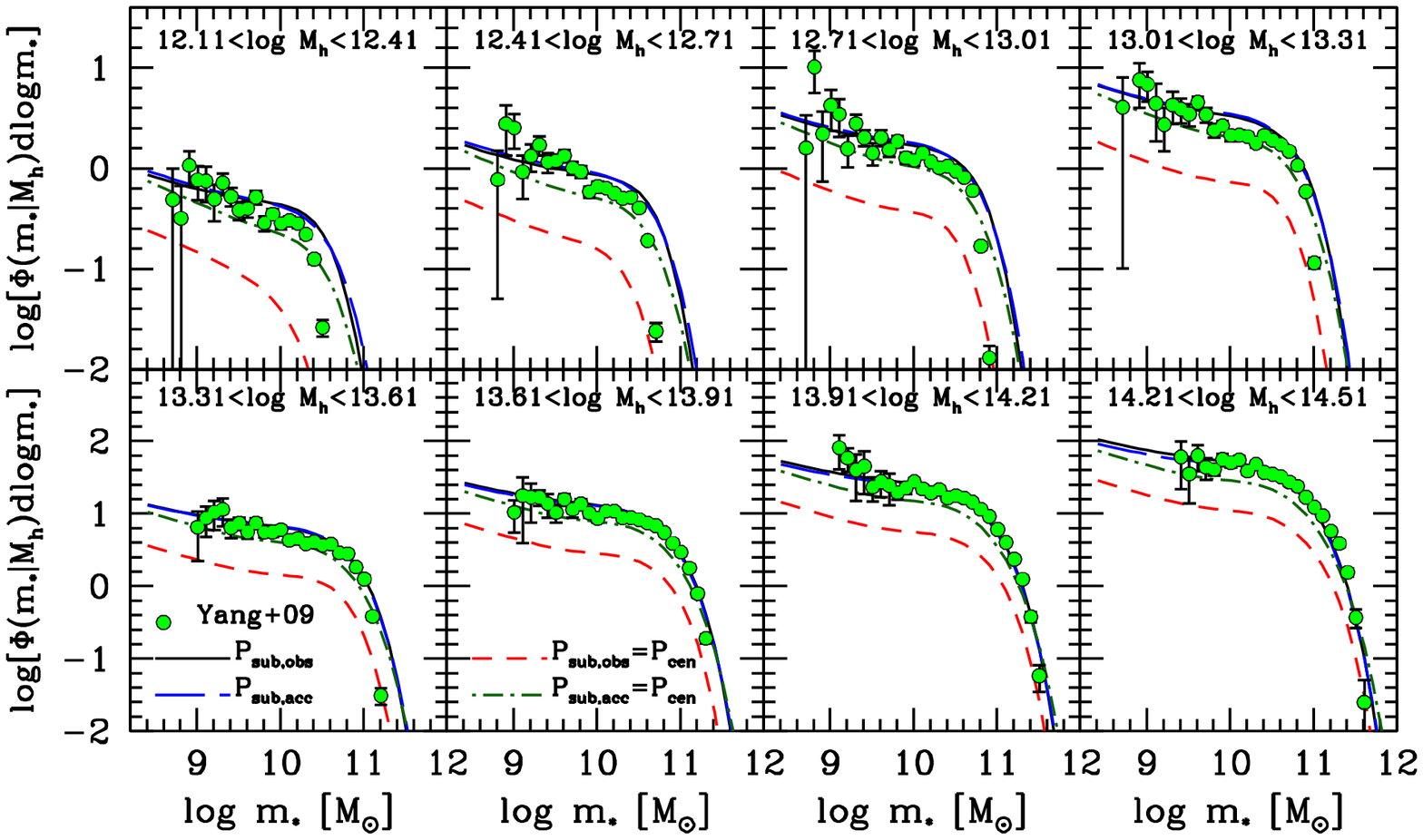}
\caption{Density--weighted average satellite $\cmf$ in eight halo mass intervals.
 Red short-dashed and green short-dashed-dot lines are for the 
 cases when the central \shmr\ and the \ssmr\ were assumed to be
equal and  \mobs\ and \macc\ were used, respectively.  The black solid and 
blue long-dashed lines are again for \mobs\ and \macc, respectively,
but in the case the central and satellite mass functions were independently
constrained by means of our extended \amt\ (they overlap most of time). 
Filled circles with error bars show the $\cmf$s inferred from observations by 
YMB09. Note that their halo masses were converted to match our virial definition.  }
\label{csmf}
\end{figure*}

From the approach described in Section 2, another statistical quantity
that deserves to be subject of study is the satellite \cmf\
(Eq.~\ref{cmfsat}).  We calculate the mean halo--density--weighted
\cmf\ at the $[M_{\rm h_1},M_{\rm h_2}]$ bin as:
\begin{equation}
\langle\Phi_s\rangle=\frac{\int_{M_{\rm h_1}}^{M_{\rm h_2}}\phis\phih(\mh) d\mh}
{\int_{M_{\rm h_1}}^{M_{\rm h_2}}\phih(\mh) d\mh}.
\end{equation}
This quantitiy has been inferred from observations by
\citetalias{YMB09}, again using their SDSS galaxy catalog (filled
circles with error bars in Fig.~\ref{csmf}).

First, we consider again the case assuming $\mss(\msub)=\ms(\mh)$.
When \msub\ is defined at the observation time, the resulting $\cmf$s
are lower than the \citetalias{YMB09} $\cmf$s by a factor of $\sim5$
in the power-law regime (roughly the same factor by which \phigp\ is
lower than the \citetalias{YMB09} observed satellite
\gsmf). Similarly, when \msub\ is defined at the accretion time, the
predicted $\cmf$s in the power-law regime are below the
\citetalias{YMB09} $\cmf$s by nearly the same factor, $\sim 2$, that
\phitwacc\ lies below the satellite \gsmf.  The normalization of the
$\cmf$ increases faster with \mh\ when \mobs\ is used instead of
\macc. This is because we allow the \mobs\ \scmf\ normalization to
vary with host halo mass, while the \macc\ \scmf\ normalization is
independent of host halo mass.

The black continuous (\mobs) and blue long-dashed (\macc) lines in
Fig.~\ref{csmf} (almost indistinguishable one from other) are the 
predictions when \Psub\ has been constrained
by means of the observed satellite \gsmf. The agreement of the
predicted satellite $\cmf$'s, for both \mobs\ and \macc\, with the
\citetalias{YMB09} $\cmf$'s is now remarkable at all halo mass bins
for low/intermedium stellar masses. Although, as above, the
normalization of the $\cmf$'s increases faster when $\msub=\mobs$ than
when $\msub=\macc$, the differences between both cases at any mass are
less than 0.05 dex, within the error bars of the observational data.

Despite the overall agreement, for halo mass bins lower than $\sim
10^{13}$ \msun, the number of massive satellite galaxies is
overestimated, specially at the lowest \mh\ bins. A possible reason for this
is the assumption that the scatter in \Psub\ is constant while in
reality it could depend on \mh\ as well as on \msub. However, the
probability of finding massive satellite galaxies in halos less
massive than $\sim 10^{13}$ \msun\ is low and they do not contribute
significantly to the mean total density of satellite
galaxies. Therefore, this assumption does not change our conclusions,
see also \S 4.
 
Our analysis shows that {\it assuming $\Psub=\Pcen$ the resulting
  satellite $\cmf$s are not consistent with observations}. Instead,
when \Psub\ is independently constrained using the observed satellite
\gsmf, there is a clear agreement, no matter what definition of
\msub\ was employed for the \scmf.

\subsection{Abundance matching and clustering}

It has been noted in the literature that the average (total) \shmr\ obtained  
with the standard \amt\ is consistent with the observed spatial clustering of galaxies
\citep{Conroy+2006,Moster+2010}. We will test now whether this is the case 
for the mass relations of central and satellite galaxies obtained here with our extended 
\amt. We will compute the galaxy projected correlation 
function by means of a HOD model for each of the mass relation obtained in \S 3.1.

A HOD model is a statistical tool mainly used to describe the
clustering of galaxies (e.g.,
\citealp{Berlind-Weinberg02,Cooray-Sheth03,Yang+03,Zehavi05,Zehavi11,
  Leauthaud+2011a, Leauthaud+2011b,Yang2011}, and more references
therein).
In contrast to the \amt, which is a quasi empirical tool, a HOD employs
modeling motivated by results of cosmological $N$-body
\citep[e.g.,][]{Kravtsov+04} and hydrodynamical
\citep[e.g.,][]{Zheng05} simulations.

In short, a HOD model describes the probability that a halo of mass
\mh\ hosts a number of $N$ galaxies with stellar masses greater than
\ms.  Once the occupational numbers are defined, the two-point
correlation function can be computed assuming that the total mean
number of galaxy pairs is the contribution of all pairs coming from
galaxies in the same halo (one-halo term) and pairs from different
halos (two-halo term). For a detailed description for the HOD model 
we employ here, see Appendix A.

\begin{figure*}
\vspace*{-200pt}
\includegraphics[height=6.5in,width=6.5in]{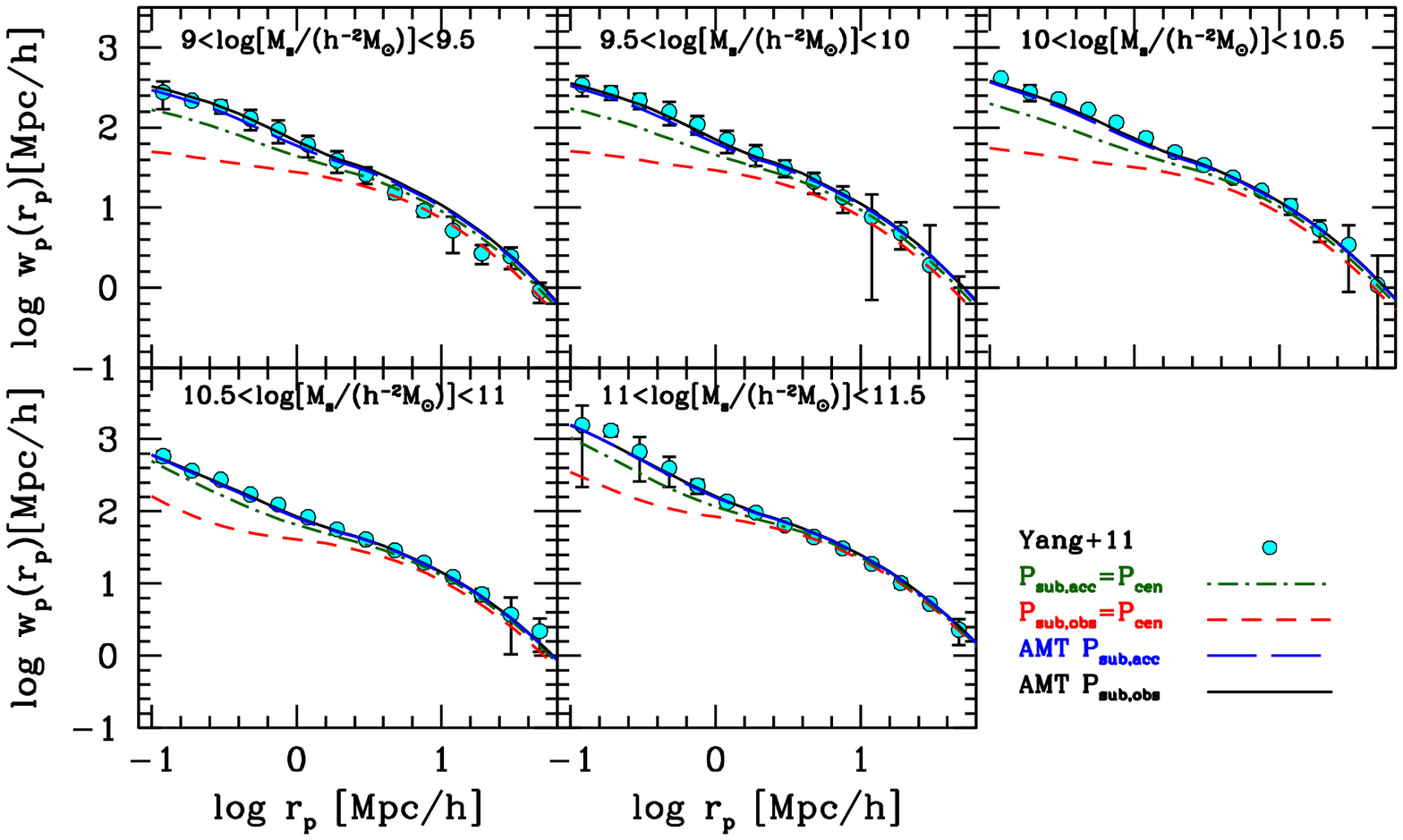}
\caption{ Projected two--point correlation functions of galaxies in five
stellar mass intervals obtained with the HOD model using different
galaxy-halo mass relations obtained with our \amt. Gray short-dashed 
and green short-dashed-dot lines are for the 
 cases when the central \shmr\ and the \ssmr\ were assumed to be
equal and \mobs\ and \macc\ were used, respectively.  The black solid and 
blue long-dashed lines are again for \mobs\ and \macc, respectively,
but in the case the central and satellite mass functions were independently
constrained (they overlap most of time).  The observed projected correlation functions 
reported in Yang  et al.\ (2011) are shown by filled circles with error bars.  }
\label{pcf}
\end{figure*}

First, consider the case when $\Psub=\Pcen$. The short-dashed curves
in Fig. \ref{pcf} show the projected correlation functions in five
stellar mass bins for the case the \mobs\ \scmf\ was used.  The
\citet{Yang2011} galaxy projected correlation functions from the DR7
SDSS are plotted as filled circles with error bars.  The resulting
correlation functions are clearly below observations, mainly in the
one-halo term.  This is because using \mobs\ underestimates the
satellite \gsmf\ and \cmf, resulting in a strong deficit of satellite
galaxies. Observe that if $N_s\sim0$, then $N\sim N_c$ and therefore,
$b_g(\ms)\sim\langle b(\mh) \rangle$ where $\langle b(\mh) \rangle$ is
the mean weighted halo bias function, see Eq.~\ref{bias_gal}.

 When using the \scmf\ for \macc\ instead of \mobs\ (dot-short-dashed
curves), the agreement with the observed correlation functions is better,
though at scales where the one-halo term dominates, the predictions are still below
observations. This is, again, because the  satellite \gsmf\ and \cmf\ are
underestimated in this case. We remark that using the (average or total) \shmr\
obtained with the standard AMT in the HOD model by  matching the total \gsmf\ 
to the total halo+subhalo mass function (in the case of \macc), leads to excellent 
agreement with the observed correlation functions, a result that is well known.  
However in this case the \ssmr\ is not constrained, instead it is 
implicitly assumed to be equal to the central
\shmr\ (for \macc). With our extended \amt, we can explicitly separate 
both mass relations.
When they are assumed to be equal and the central \shmr\ is constrained with
the central \gsmf, then we obtain the predictions already shown, in particular
the correlation functions. The fact that the predicted correlation functions,
when \macc\ is used, are close to those predicted in the standard \amt\ (and to
the observed ones) is because the central and average \shmr\ are indeed close,
as we discussed in \S\S 3.1, see Fig.~\ref{msmh}.

Thus, under the assumption that $\Psub=\Pcen$, the observed clustering
of galaxies is better reproduced when the subhalo mass in abundance
matching is defined as \macc\ rather than \mobs. Nevertheless, even in
the former case, the agreement with observations is only marginal.

We now turn the analysis to the cases where the \ssmr\ is not assumed
to be equal to the central \shmr.  The black solid and blue
long-dashed lines in Fig. \ref{pcf} show the predicted correlation
functions in the cases where either \msub\ or \macc\ were used.  Both
cases lead to very similar results and agree very well with
observations.

Therefore, {\it the HOD model combined with the central and satellite mass 
relations independently constrained with the extended \amt, is able to reproduce 
the observed correlation functions, no matter if \mobs\ or \macc\ are used}. This
successful prediction is a consequence of the good agreement obtained
between our predicted satellite $\cmf$s and those inferred from
observations (\S\S 3.2 and Fig. 3).

\section{Summary and Discussion}

In this paper, we extend the \amt\ in order to constrain both the
central stellar--halo and the satellite--subhalo mass relations
separately, using as an input (i) the distinct halo and subhalo
mass functions, and (ii) the observed central and
satellite $\gsmf$s. Our formalism, by construction, predicts also the
satellite $\cmf$s as a function of host halo mass, and when applied to
a HOD model, allows to predict the spatial correlation functions.  We
present results for the cases when the \ssmr\ is assumed to be equal
to the central \shmr, \Psub=\Pcen, and when both mass relations are
constrained independently (i.e., it is not assumed that \Psub=\Pcen).
All our analysis is carried out for subhalo masses defined at
accretion time, \macc, and at the observed time (present day),
\mobs. The main results and conclusions are as follows:

\begin{itemize}
\item Assuming that the mass relation between satellites and subhalos
  is identical to the mass relation between centrals and distinct
  halos (including their intrinsic scatters), \Psub=\Pcen, the predicted satellite \gsmf, $\cmf$s and
  projected two--point correlation functions lie below those obtained
  from observations for both definitions of \msub, though the
  disagreements are small when \msub=\macc\ (Figs. \ref{gsmf},
  \ref{csmf}, \ref{pcf}).  We conclude that assuming \Psub=\Pcen\
  leads to predictions in disagreement with observations, specially
  when \mobs\ is used.

\item When the \ssmr\ is no longer assumed to be equal to the central \shmr\, and
  instead is constrained by means of the observed satellite \gsmf,
  the predicted satellite $\cmf$s and projected correlation functions
  agree in general with observations, both for \mobs\ and \macc.
  However, for halo masses lower than $\sim 10^{13}$ \msun, the number
  of very massive (rare) satellites is over-predicted.

\item The resulting \msub--\mss\ relations when using either
    $\mobs$ or $\macc$ are quite different from each other, and in
  each case are different from the central \shmr\
  (Fig. \ref{msmh}). For a given stellar mass, the satellite subhalo
  mass is smaller than central halo mass, and the mass difference is
  increasing the lower the mass is. These differences are dramatic
  when \mobs\ is used.

\item Our density-weighted average (centrals + satellites) $\shmr$s are
  close to the central \shmr\ when either \mobs\ or \macc\ is used
  (central galaxies dominate in the YMB09 catalog).  Such an average
  \shmr\ coincides conceptually with the one inferred from matching
  the total (centrals+ satellites) cumulative \gsmf\ and the halo +
  subhalo cumulative mass function (standard \amt).
\end{itemize}

\subsection{On the inference of the \ssmr\ and its implications for
  the average mass relation}

The conclusions listed above can be well understood by examining the
basic ideas behind the extended AMT, as we show in Section 2.2. Essentially,
matching abundances of satellite galaxies to subhalos is equivalent to
matching their corresponding occupational numbers, that is:
\begin{equation}
\Ns=\Nsub.
\label{amt_cmf}
\end{equation}
The opposite is also true: matching their corresponding occupational
numbers is equivalent to matching their abundances.
This is an important result because it shows 
that once \Psub\ (and \Pcen) is properly constrained, we 
will obtain the correct conditional mass functions
and consequently the correct spatial clustering for galaxies.

The above means that there is a unique \msub(\mss) relationship for
each definition of \msub, which depends solely on \Nsub. Because of
this uniqueness, it follows that the \mobs(\mss) and \macc(\mss)
relations should be different, and any incorrect assumption on each
one of these relations will lead to inconsistencies in the conditional
mass functions and spatial clustering of galaxies, as for example
those that we have found here when \Psub\ was assumed to be equal to
\Pcen.  Under this assumption, when \macc\ was used, the
inconsistencies were actually small. This is because in this case the
"incorrect" assumption for the \ssmr\ is actually not too far from the
"correct" result obtained when \Psacc\ is independently constrained
(compare dot-dashed and solid green curves in Fig. \ref{msmh}),
contrary to what happens when \mobs\ is used.

It is important to remark that in the standard \amt, only the average \shmr\
is constrained (using the total \gsmf), leaving unconstrained the \ssmr, something 
that on its own introduces a large uncertainty in the average \shmr\ \citep[see][]{Neistein11}. 
We have shown that such average \shmr\ is conceptually equal to the density--weighted average 
mass relation obtained here from the observationally constrained central \shmr\ and \ssmr. 
Therefore, our resulting average mass relation is expected to be less uncertain than
previous determinations. On the other hand, this average mass relation is expected 
to be close to the central \shmr\ because most of the galaxies in the used observational 
catalog are centrals.  

We conclude that in order to properly infer the \ssmr\ and the central 
\shmr\ at the same time, and this way reduce the uncertainty in the average \shmr, 
more observational constraints than the total \gsmf\ are necessary.
The most obvious and direct is the \gsmf\ decomposed into central and satellite
galaxies, something that was provided by YMB09. 
However, observe that, according to eq. ~(\ref{amt_cmf}), 
the satellite \cmf s or the clustering of galaxies, 
 modulo the observational errors, provide observational constraints that lead 
to similar inferences of the \ssmr, because of the uniqueness of this  
relation for a given well defined \Nsub\ (see above).

Finally, we note that obtaining the \ssmr\ for the subhalo mass
defined at the accretion time introduces uncertainties due to our
ignorance about evolutionary processes of the stellar mass since accretion . 
This does not happen when
the \ssmr\ is obtained for both the satellite and subhalo masses defined
at the same epoch, for instance the present time.
When matching abundances for the \macc\ case, the fact that (1) \macc\ is itself a
cumulative distribution of all objects accreted over a period of time,
and that (2) \mss\ may have changed between accretion and observation,
are not taken into account. In other words, it is implicitly assumed that the
satellite stellar mass stops evolving soon after accretion. 
In reality the situation is actually quite complex in the sense that, depending on 
the accretion time and the orbit of the satellites, the evolution of their stellar 
masses is diverse, with some of them early quenched and others actively 
evolving, perhaps in some cases as the central ones of the same mass
\citep[see e.g.,][for semi-empirical inferences on such  a complexity
of galaxy evolution in groups]{Wetzel+2011}.
This diversity introduces an intrinsic uncertainty on the results. 
Such an uncertainty might be accounted for the probability distribution functions:
$P(\mss|m_{*,\rm acc},z)$, which 
gives the probability that a satellite accreted at epoch $z$ evolves, on average, to
the observed satellite \mss, and $P(m_{*,\rm acc}|\macc,z)$, which gives the probability 
that a subhalo $\macc$ hosts a galaxy of mass $m_{*,\rm acc}$
at the time of accretion. Now, the satellite \cmf\ (Eq.~\ref{cmfsat}) can be written as
\citep{Mo+2010}:
\begin{eqnarray}
\phis=\int\int\int P(\mss|m_{*,\rm acc},z)P(m_{*,\rm acc}|\macc,z)\times & \nonumber \\
\Phi(\macc|\mh,z)dm_{*,\rm acc} d\macc dz.& \nonumber \\
\label{fisat-macc}
\end{eqnarray}
Note that in our analysis in \S\S 3.1,
we implicitly assume that the stellar mass of satellite galaxies
does not change once they become satellites, i.e. $P(\mss|m_{*,\rm acc},z)
=\delta(\mss-m_{*,\rm acc},z)$, and that $P(m_{*,\rm acc}|\macc,z)$
is independent of redshift. Thus, the application of the AMT to infer
the satellite \cmf\ and the \mss--\msub\ relation for subhalo mass defined at 
its accretion time formally requires more observational constraints at higher 
redshifts. This is a problem already faced by previous authors
\citep[e.g.,][]{Yang2011}.

The above is not the only way to formally write the satellite
\cmf; it can be written in a way that instead of implying knowledge of the
change of \mss\ from accretion to observation, implies 
just knowledge on the change of the \scmf\ between these two epochs. 
Let us consider the distribution function, \Pacc, giving the
probability that halos accreted at epoch $z$ evolve, on average, to
the observed (present-day) subhalos \mobs, and the probability distribution function of these
subhalos of hosting a galaxy of mass \mss, \Psacc. In this
case, the satellite \cmf\ (Eq.~\ref{cmfsat}) is written as
\begin{eqnarray}
\phis=\int\int\int\Psacc\Pacc\times & \nonumber \\
\Phi(\macc|\mh,z)d\mobs d\macc dz,& \nonumber \\
\end{eqnarray}
and therefore the satellite \gsmf, Eq.~\ref{phisat}, is given by
\begin{eqnarray}
\phisat(\mss)=\int\int\int\Psacc\Pacc\times & \nonumber \\
\densub(\macc,z)d\mobs d\macc dz.& \nonumber \\
\label{fisat-obs}
\end{eqnarray}
Since the \macc\ \shmf\ would evolve into the \mobs\ \shmf, we write
\begin{equation}
\densub(\mobs)=\int\int\Pacc\densub(\macc,z)d\macc dz.
\label{mob-macc}
\end{equation}
This last equation is the abundance matching of accreted subhalos to
present-day subhalos. Therefore,
\begin{equation}
\phisat(\mss)=\int\Pobs\densub(\mobs)d\mobs.
\label{fisat-mobs}
\end{equation}
This equation is nothing but abundance matching satellite galaxies to
subhalos \emph{at the time they are observed}. Hence, the reason that 
the satellite \gsmf\ matches the \shmf\ in a more
direct way for subhalo masses defined at the observation time
(eq.~\ref{fisat-mobs}) than at the accretion time
(eq.~\ref{fisat-macc}), is that in the latter case the unknown 
$P(\mss|m_{*,\rm acc},z)$ and $P(m_{*,\rm acc}|\macc,z)$
"evolutionary" functions have to be introduced.
However, we acknowledge that for the former case,
our ignorance on the scatter around the \ssmr\ is a also potential
source of uncertainty. All our calculations are under the assumption
that this scatter is the same as the scatter of the central \shmr. 
In any case, even if these scatters are different, note that including
scatter affects the stellar-to-(sub)halo mass relation only at its 
high-mass end, where on average satellites are expected to be 
accreted recently, hence their \ssmr\ and scatter are yet similar to those 
of centrals/halos. 

\subsection{On the intrinsic scatter in the \ssmr\ }

A possible source of systematic errors in our analysis is the
assumption that the intrinsic scatter around the \ssmr, $\sigma_s$ is
constant and equal to the scatter around the central \shmr.  To probe
the impact of this assumption we repeated all our analysis but this
time assuming $\sigma_s=0$.  When comparing the results using
$\sigma_s=0$ to those obtained based on $\sigma_s=0.173$ dex, we find
that they are consistent with each other, and therefore with the
satellite \cmf s and with the galaxy spatial clustering measured from
the \citetalias{YMB09} catalog. In more detail, we find that the
resulting $\cmf$'s reproduce observations for $\sigma_s=0$ slightly
better than for $\sigma_s=0.173$ dex, especially at the massive end.
This is because when the intrinsic scatter is not taken into account
($\sigma_s=0$), the shape of the \ssmr s steepens at the massive end
(see also \citealp{Behroozi+2010}).  Consequently, for a given \mss,
the subhalo mass is larger, and the abundances of larger (sub)halos is
lower in general than those of smaller halos.  Therefore, the number
density of satellites at the massive end is lower. However, the
projected correlation functions remain almost the same because the
probability of finding a massive satellite galaxy in host halos less
massive than $\log\mh\sim13$ is very low. They do not contribute
significantly to the mean total density of galaxies. Although better
models are needed in order to give a realistic form for $\sigma_s$,
our main conclusions seem to be robust to variations in the adopted
value for $\sigma_s$.

\subsection{Implications for satellite/subhalo evolution}

The local \ssmr\ obtained for both definitions of the subhalo mass,
\macc\ and \mobs, are such that at halos masses smaller than
$2-10\times 10^{13}$ \msun\ and at a given galaxy stellar mass, the
corresponding subhalo mass is smaller on average than the halo mass of
centrals (Fig. \ref{msmh}). This difference increases the smaller the
mass is, and much more for the subhalo mass defined at the observed
time (present-day).  In the case of \macc, the differences might be
because the halo mass at the epoch it became a subhalo (accretion
time) is smaller than its present-day counterpart at a given stellar
mass and/or because the satellite stellar mass increased faster than
the central one for a given halo mass. In fact, it is difficult to
make any inference in this case because the abundance matching is
between \emph{local} galaxies and (sub)halos at \emph{different past}
epochs.  In any case, the fact that the inferred mass relations for
satellites and centrals when \macc\ is used are not too different,
suggests that the central galaxy--distinct halo mass relation does not
change too much with time, at least since the epoch at which most of
the subhalos were accreted.

When \mobs\ is used, both abundances of satellites and subhalos are
matched at the same epoch, the observation (present-day) time.  In
this case the strong difference between the satellite and central mass
relations can be interpreted mainly as the result of subhalo mass loss
due to tidal stripping. Besides, the smaller the subhalo, the larger
is the mass loss on average. Probably, the different evolution in
stellar mass between central and satellite galaxies could also play a
role for the differences but not as significant a role the one related
to halo and subhalo mass evolution. 

 From Fig. \ref{msmh} one sees
that for a given \mss, the \mobs--to--\macc\ ratio is 0.35--0.40 for
the smallest masses up to $\mss\sim 2\times10^{11}$ \msun.  At larger
masses, this ratio rapidly tends to 1. Therefore, the subhalos of
satellites galaxies less massive than $\mss\sim 2\times10^{11}$ \msun\
have lost, on average (for all host halo masses\footnote{The dependence of the satellite subhalo mass loss on host halo
mass will be explored elsewhere.}), 65--60\% of their
masses since they were accreted. It should be noted that
this is a rough approximation and the evolution of the stellar mass 
since the satellite was accreted should be taken into account, see \S4.1. 
This above result shows us that the galaxy-(sub)halo connection for satellite
galaxies is far from direct; present-day satellites of masses
$\mss\sim 7\times 10^8$ \msun\ and larger have formed in subhalos that
at the time they were accreted onto galaxy sized halos were on average
a factor 2.5--3 larger than at the present epoch. This has severe
implications for studies aimed to constrain the \lcdm\ scenario
at the level of subhalo/satellite distributions. 

For example, it has been discussed that seeding the subhalos in simulations
of MW-like halos by using an extrapolation to low masses of the stellar--halo
mass relation obtained by means of the \amt, predicts a MW dwarf spheroidal (dSph)
luminosity function in agreement with the observed one. However, the 
circular velocities at the maximum (or the masses at the infall) of 
the subhalos associated to the dSphs result significantly
larger than inferences from observed kinematics \citep[][]{Boylan-Kolchin+2011a}. 

In the right panel of Fig.~\ref{msmh} we have plotted the extrapolation to
low masses of our \ssmr s, both for subhalo masses defined at the present 
day (red line) and at the infall time (blue line). The observational points
in the panel are for MW satellites, which subhalo masses were estimated
at their truncation radii. Thus, if we assume that these masses are
roughly equal to the present-day subhalo masses in the \lcdm\ simulations, then
the simulated subhalo masses, \mobs, are up to $\approx 10-30$
times larger than those associated to dSphs. If the comparison
is done with the extrapolation of the average (or central) \shmr, then
the differences increases by a factor of $\sim 3$ more 
\citep[see also Fig. 7 in][]{Boylan-Kolchin+2011a}.

Our extrapolated results show that the discrepancy in subhalo mass 
between MW bright dSphs and \lcdm\ simulations is smaller than previously 
reported but is still significant. Note that for the extrapolation, we have used
the same slope of the \citetalias{YMB09} satellite \gsmf\ at the low mass end,
$\alpha=-1.25$ (Fig. \ref{gsmf}). If this slope steepens for smaller masses, 
for example to a value of $\alpha=-1.6$, then our extended \amt\ predict
the \mobs\ \ssmr\ plotted as the gray dotted-dashed curve in Fig. ~\ref{msmh},
which is already consistent with the dynamical estimates. 

The \gsmf\ at low masses may be significantly incomplete because of
missing low-surface brightness galaxies. 
By taking into account the bivariate distribution of stellar mass versus
surface brightness, \citet{Baldry+2008} have found evidence for an upturn in the
faint-end \gsmf\ slope ($\alpha\approx-1.6$) for a subsample of field
SDSS galaxies.  More recently, using the GAMA survey, a slope of 
$\alpha\approx-1.47$ has been reported \citep{Baldry+2011}. Steep faint-end
slopes have been also found at higher redshifts. For instance,
using the COSMOS field, \citet{Drory+09} have measured 
slopes of $\alpha\sim -1.7$ at all redshifts $z\le 1$
There are also pieces of evidence that the faint-end slope
of the \gsmf\ (or luminosity function) changes with the environment:
in clusters of galaxies it steepens significantly 
\citep[for a discussion see][and the references therein]{Baldry+2008}.
The cluster \gsmf\ is actually related to the satellite \gsmf, through
the satellite \cmf. 

We conclude that using a correct \amt\ for connecting satellite
galaxies to their present-day subhalos and assuming a steep faint-end slope
in the satellite \gsmf\ ($\alpha\sim-1.6$), the predicted 
\ssmr\ for the \lcdm\ cosmogony would be consistent 
with the dynamics of MW satellites. 

\acknowledgments
We thank Alexie Leathaud,  Ramin Skibba, Surhud More, 
Xiaohu Yang, and the anonymous Referee  for useful comments, 
and suggestions.
A. R-P acknowledges a graduate student 
fellowship provided by CONACyT.
N. D. and V. A. acknowledge to CONACyT grant 
128556 (Ciencia B\'asica). 
V.A acknowledges PAPIIT-UNAM grant IN114509.
We are grateful to X. Yang for providing us
in electronic form their data for the $\cmf$s. 

\bibliographystyle{mn2efix.bst}
\bibliography{blib}

 \appendix
  
 \section{The spatial clustering of galaxies
 in the HOD model}
 
Here we review the main ideas used to infer the spatial clustering of
galaxies based on a HOD model. We assume that the most massive galaxy
in terms of stellar mass within a halo of mass \mh\ is its central
galaxy. Consequently the remaining galaxies are all satellites. We let
them follow the mass density profile of the host halo.  We denote the
cumulative number of central and satellite galaxies with stellar
masses greater than \ms\ as $N_c$ and $N_s$, respectively.

The two point correlation function is decomposed into two terms,
\begin{equation}
1+\xigg=[1+\xiggh]+[1+\xigghh],
\end{equation}
with $1+\xiggh$ describing galaxy pairs within the same halo (the
one-halo term), and $1+\xigghh$ describing the correlation between
galaxies occupying different halos (the two-halo term).

We compute the one-halo term as
\begin{eqnarray}
1+\xiggh=\frac{1}{2\pi r^2n_g^2}\int_{0}^{\infty}\frac{\langle N(N-1)\rangle}{2}\lambda(r)\phih(\mh) d\mh,
\end{eqnarray}
for pairs separated by a distance $r\pm dr/2$. 
Here $\langle N(N-1)\rangle/2$ is the total mean number 
of galaxy pairs within halos \mh\ following a pair distribution 
$\lambda(r)dr$, and a mean number density \ng(\ms).
The contribution to the total mean number of galaxy pairs from 
central-satellite pairs and satellite-satellite pairs is
\begin{eqnarray}
\frac{\langle N(N-1)\rangle}{2}\lambda(r)dr=\langle N_c\rangle\Nsat\lambda_{c,s}(r)dr& & \nonumber\\
+\frac{\langle N_s(N_s-1)\rangle}{2}\lambda_{s,s}(r)dr.
\label{pair}
\end{eqnarray}
As commonly assumed in HOD models, the number of central-satellite
pairs follow the normalized mass density halo profile, taken to be of
\citet{NFW97} shape.  The number of satellite-satellite pairs is then
related to the normalized density profile convolved with itself.

Halo profiles are defined in terms of the total halo mass and the
concentration parameter. We use the relation between concentration
parameter $c_{\rm NFW}$ and halo mass obtained by \citet{Munoz11} from
fits to $N$-body simulations.
 
Based on results of high-resolution $N$-body \citep{Kravtsov+04} and
hydrodynamic simulations of galaxy formation \citep{Zheng05}, we model
the second moment of satellite galaxies, $\langle N_s(N_s-1)\rangle$,
as a Poisson distribution with mean $\Nsat^2=\langle
N_s(N_s-1)\rangle$.

We compute the two-halo term as
\begin{equation}
\xigghh=b_g^2\zeta^2(r)\xi_m(r),
\end{equation}
where $\xi_m(r)$ is the non-linear matter correlation function \citep{Smith03}, 
$\zeta(r)$ is the scale dependence
of dark matter halo bias \citep{Tinker05}, and 
\begin{equation}
b_g= \frac{1}{\ng}\int_{0}^{\infty}b(\mh)\NG\phih(\mh) d\mh,
\label{bias_gal}
\end{equation}
is the galaxy bias with $b(\mh)$ the halo bias function \citep{Sheth03}.

Once we have calculated \xigg, we relate it to the projected
correlation function, \wp, integration over the line of sight,
\begin{equation}
\wp=2\int_{0}^{\infty}\xi_{\rm gg}(\sqrt{r_{\rm p}^2+x^2})dx.
\end{equation}

\end{document}